\documentclass[natbib209]{emulateapj}
\shortauthors{Miesch and Dikpati}
\shorttitle{3D BABCOCK-LEIGHTON DYNAMO}

\usepackage{epsfig}

\newcommand{\pd}{\partial}

\newcommand{\del}{\mbox{\boldmath $\nabla$}}
\newcommand{\curl}{\mbox{\boldmath $\nabla \times$}}

\newcommand{\cross}{\mbox{\boldmath $\times$}}

\newcommand{\vv}{{\bf v}}
\newcommand{\BB}{{\bf B}}

\begin{document}

\title{A 3D BABCOCK-LEIGHTON SOLAR DYNAMO MODEL}

\author{Mark S. Miesch and Mausumi Dikpati}

\affil{High Altitude Observatory, National Center for Atmospheric
Research, 3080 Center Green, Boulder, CO 80307-3000.}

\email{miesch@ucar.edu}

\begin{abstract}

We present a 3D kinematic solar dynamo model in which poloidal field
is generated by the emergence and dispersal of tilted sunspot pairs
(more generally Bipolar Magnetic Regions, or BMRs).  The axisymmetric
component of this model functions similarly to previous 2D
Babcock-Leighton (BL) dynamo models that employ a double-ring
prescription for poloidal field generation but we generalize this
prescription into a 3D flux emergence algorithm that places BMRs on
the surface in response to the dynamo-generated toroidal field.  In
this way, the model can be regarded as a unification of BL dynamo
models (2D in radius/latitude) and surface flux transport models (2D
in latitude/longitude) into a more self-consistent framework that
captures the full 3D structure of the evolving magnetic field.  The
model reproduces some basic features of the solar cycle including an
11-yr periodicity, equatorward migration of toroidal flux in the deep
convection zone, and poleward propagation of poloidal flux at the
surface.  The poleward-propagating surface flux originates as trailing
flux in BMRs, migrates poleward in multiple non-axisymmetric streams
(made axisymmetric by differential rotation and turbulent diffusion),
and eventually reverses the polar field, thus sustaining the dynamo.
In this letter we briefly describe the model, initial results, and
future plans.
\end{abstract}


\section{INTRODUCTION}\label{sec:intro}

\citet{babco61} was the first to describe how the emergence of
toroidal magnetic flux through the solar surface and the subsequent
evolution of that flux can produce a large-scale poloidal magnetic
field.  Furthermore, he argued that this process, together with the
generation of toroidal field by differential rotation (the
$\Omega$-effect) gives rise to the 11-yr solar activity cycle.
Later work beginning with \citet{leigh64,leigh69} fleshed out
Babcock's vision and transformed it into viable numerical dynamo
models of the solar cycle. 

Though many alternative solar dynamo models have been proposed, 
the Babcock-Leighton (BL) paradigm has remained compelling because 
it is firmly grounded in solar observations and provides a robust
mechanism for producing cyclic dynamo activity 
\citep[see reviews by][]{dikpa09,charb10}.  One of the major
milestones in model development occurred in the 1990s when 
meridional circulation was included and was shown to play
a crucial role in regulating the cycle period and other 
cycle features such as the poleward drift of photospheric 
flux and the phasing of polar field reversals 
\citep{wang91,choud95,dikpa99b}.  In recognition
of the importance of flux transport by meridional circulation,
these new BL models were christened Flux-Transport (FT) dynamo
models and remain popular today.

Though they ostensibly rely on flux emergence and evolution in order
to operate, most early BL/FT models did not explicitly include
sunspots. instead, the generation of poloidal field through the
BL mechanism was represented as an idealized axisymmetric source 
term in the poloidal component of the magnetohydrodynamic (MHD) induction 
equation. This BL source term is often nonlocal in the sense that it is 
confined to the surface layers, but its amplitude is proportional to the 
strength of the toroidal field near the bottom of the convection zone (CZ).  
The 2D (axisymmetric) MHD induction equation is then solved to follow the 
evolution of kinematic, axisymmetric (longitudinally-averaged) mean fields.  

Another milestone in model development was to replace the non-local 
$\alpha$-effect with a more phenomenological representation of tilted
sunspot pairs.  This was originally done in an axisymmetric context
through Durney's (1997\nocite{durne97}) double-ring algorithm which 
represents a tilted sunspot pair as two overlapping toroidal rings
with opposite polarity.   This algorithm was later extended and
implemented into 2D BL/FT dynamo models by \cite{nandy01},
\cite{munoz10} and \cite{guerr12}.

A more sophisticated 3D flux emergence algorithm was recently 
presented by Yeates \& Munoz-Jaramillo (2013; hereafter 
YM13\nocite{yeate13}).  To our knowledge, this is the first 
use of a fully 3D Babcock-Leighton source term.
In their model, YM13 model flux emergence through an 
imposed helical flow that lifts and twists the dynamo-generated
toroidal field such that it emerges through the surface 
and then evolves according to the action of turbulent 
diffusion and mean fields.   

Here we explore an alternative approach to a 3D kinematic
BL/FT model.  Rather than imposing a flow to advect the 
magnetic field upward as in YM13, we place a spot pair (or, more
generally a Bipolar Magnetic Region, or BMR; cf. YM13)
confined to the surface layers above the position where 
the subsurface toroidal flux peaks.  Since the mean-field
component of the 3D induction equation is equivalent to 
a corresponding double-ring algorithm, this approach makes
closer contact with previous 2D (latitude-radius) BL/FT
dynamo models.  Furthermore, Since the emergent field is 
confined to the surface layers, it makes closer contact
with a seperate class of models known as surface flux transport
(SFT) models that follow the 2D (latitude/longitude) evolution of
emergent flux in the solar photosphere subject to mean flows and
turbulent diffusion \citep{leigh64,wang91,schri01,bauma06}.

In summary, our model is a unification of BL/FT dynamo models
and SFT models.  Though it is not the first such unification
(see Munoz-Jaramillo et al. 2010, YM13\nocite{munoz10,yeate13}),
it is a promising approach that we
intend to pursue in the future to study the 3D evolution of the cyclic
solar magnetic field and its coupling to the corona and heliosphere.
We describe the basic model components in \S\ref{sec:model} and the
flux emergence algorithm in \S\ref{sec:spotmaker}. We then present
illustrative results, conclusions, and future plans in
\S\ref{sec:results}.

\section{DEVELOPMENT OF 3D BABCOCK-LEIGHTON DYNAMO MODEL}\label{sec:model}

Building on the success of previous 2D BL/FT dynamo models, we
construct a 3D solar dynamo model by solving the MHD induction equation in the 
kinematic limit
\begin{equation}\label{eq:indy}
\frac{\pd \BB}{\pd t} = \curl \left(\vv \cross \BB - 
\eta_t \curl \BB\right)
\end{equation}
where $\eta_t(r)$ is a turbulent diffusion, and solar velocity fields are 
specified based on photospheric observations and helioseismic inversions. 
We use spherical polar coordinates ($r$,$\theta$,$\phi$) throughout.
Unlike many mean-field dynamo models, we do not include an explicit 
$\alpha$-effect.   Instead, the dynamo is sustained by the appearance 
and evolution of sunspot pairs (BMRs) which are placed on the surface in response
to the dynamo-generated field by the ``Spotmaker'' algorithm
described in \S\ref{sec:spotmaker}.

In this introductory paper, the velocity field $\vv$ is axisymmetric,
consisting only of differential rotation and meridional circulation.
In this case the evolution of the mean field $\left<\BB\right>$
(brackets denote an average over longitude, $\phi$) is independent of
modes with higher azimuthal wavenumbers ($m > 0$).  This can be
verified by averaging eq.\ (\ref{eq:indy}) over longitude.  Thus, from
the perspective of the mean fields, the Spotmaker algorithm is
equivalent to the the double-ring approach used in previous 2D BL/FT
dynamo models \citep{durne97,nandy01,munoz10,guerr12}, though details
such as the spatial profiles and temporal cadence of the spot appearances
are different.  We view this as beneficial at this stage in the model
development because it allows us to make direct contact with existing
2D BL/FT models.  Future modeling will incorporate non-axisymmetric flow
fields and nonlinear feedbacks (see \S\ref{sec:results}) which will 
break this degeneracy with 2D models but for now it provides an auspicious 
framework to build upon previous work.

The framework of the model is built upon the ASH (Anelastic Spherical
Harmonic) code described by \citet{clune99} and \cite{brun04}.  
ASH is a workhorse code that has been applied extensively to simulate
solar and stellar convection \citep[see][]{miesc05,brun10b} but
here we use it in a kinematic mode to solve only the induction
equation. The numerical method is pseudospectral, with a 
triangularly-truncated spherical harmonic decomposition in 
the horizontal dimensions and mixed semi-implicit/explicit timestepping.
This version of the ASH code uses a fourth-order finite difference 
formulation in the radial dimension.

The differential rotation and meridional circulation that comprise $\vv$ are
expressed in terms of an angular velocity profile $\Omega(\theta,r)$ and a mass
flux stream flunction $\Psi(\theta,r)$ with the same formulations and parameter 
values used in the 2D models of \citet{dikpa11}.  For the turbulent diffusion 
$\eta_t$ we use the two-step profile described by \citet{dikpa07}.

\section{FLUX EMERGENCE ALGORITHM: SPOTMAKER}\label{sec:spotmaker}

Our objective is to construct a solar dynamo model that captures both the solar
activity cycle and the observed evolution of large-scale magnetic flux
solar surface.  However, capturing the full complexity of active 
region formation and dispersal through is currently beyond the capability
of a single numerical dynamo model.  Here we use an idealized flux emergence algorithm to place spots on the solar
surface in response to the dynamo-generated toroidal field near the base of the
CZ. As mentioned in \S\ref{sec:intro} this algorithm can be regarded as a 
3D generalization of the axisymmetric double-ring algorithm of
\citet{durne97} and \citet{munoz10}.

The first step in the algorithm is to define a spot-producing toroidal 
flux near the base of the CZ as follows
\begin{equation}\label{Bhatsm}
\hat{B}_\phi(\theta,\phi,t) = \int_{r_a}^{r_b} h(r) B_\phi(r,\theta,\phi,t) dr 
\end{equation}
where $h(r) = h_0 \left(r - r_a\right)\left(r_b - r\right)$
and $h_0$ is defined such that $\int_{r_a}^{r_b} h(r) dr = 1$.
This is similar to analogous expressions used by \citet{rempe06}, but unlike 
previous models, the flux $\hat{B}_\phi(\theta,\phi,t)$ is not necessarily 
axisymmetric; longitudinal structure is permitted in the toroidal bands 
that give rise to active regions.

\begin{figure*}
\centering
\centerline{\epsfig{file=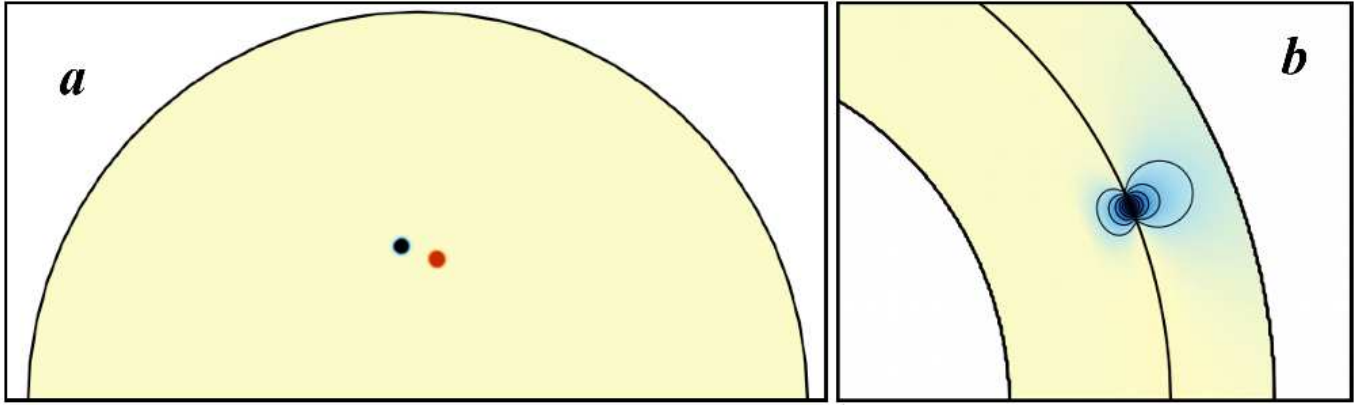,width=\linewidth}}
\caption{\small\label{fig:spot}
Structure of a spot pair generated by the flux emergence algorithm, 
Spotmaker. ($a$) orthographic projection of radial field at the solar 
surface illustrating a mid-latitude spot with a Joy's law tilt.  Blue
and red denote inward and outward field respectively. ($b$) Mean 
(axisymmetric) component of the poloidal field associated with
the tilted spot pair.  Colors and contours correspond to the
magnetic potential $\Gamma$, with blue denoting a counter-clockwise
field orientation.  Here the penetration radius $r_p = 0.95R$.
}
\end{figure*}

The next step is to suppress sunspot formation at high latitudes.
This is empirically motivated but may have a dynamical explanation
in terms of the disruption of high-latitude toroidal flux systems 
by the magneto-rotational instability \citep{parfr07}. We accomplish
this by applying a mask to $\hat{B}_\phi(\theta,\phi,t)$ such that
\begin{equation}\label{Bstar}
B^*(\theta,\phi,t) = 
\frac{2 g_0 \vert \sin\theta \cos\theta \vert}{1+\exp[-\gamma_s \theta^\prime]} 
~~ \hat{B}(\theta,\phi,t) ~~~,
\end{equation}
where $\theta^\prime = \theta - \pi/4$ in the northern hemisphere (NH) 
and $3\pi/4 - \theta$ in the southern hemisphere (SH).  Here we use $\gamma_s = 30$ and
choose the normalization $g_0$ such that the maximum value of 
the masking function is unity \citep[see][]{dikpa04b}. 

The placement of a spot pair in latitude and longitude is given by
the location where the amplitude of $B^*(\theta,\phi,t)$ is maximum.
If this occurs over a broad range of longitude (for example, from 
axisymmetric initial conditions), then a longitude is chosen at
random from those locations where $B^*(\theta,\phi,t)$ is within 
0.1\% of its peak value.

A spot is placed if the maximum amplitude of $B^*$ exceeds a 
threshold value $B_t$.  However, in order to avoid introducing
overlapping spots at every time step, a time delay is also required.
This can be loosely regarded as a dynamical adjustment time between 
flux emergence events.  Here
we use a cumulative lognormal distribution function defined as
\begin{equation}\label{cum}
{\cal C}(\Delta) = \frac{1}{2} \left[1 - \mbox{erf}\left(- 
\frac{\ln \Delta - \mu}{\sigma \sqrt{2}}\right)\right] ~~~.
\end{equation}
where $\Delta = t - t_s$ is the time lag since the last appearance
of a spot, $t_s$, and $\mu$ and $\sigma$ are parameters related
to the mean time between spots, $\tau_s$ and the mode of the
distribution, $\tau_p$ as follows:
\begin{equation}\label{lognorm1}
\sigma^2 = \frac{2}{3} ~ \left[\ln (\tau_s) - \ln (\tau_p)\right]
\mbox{\hspace{.2in} and \hspace{.2in}}
\mu = \ln{\tau_p} + \sigma^2 ~~~.
\end{equation}
Spots are placed if $\mbox{max}(B^*) > B_t$ and 
${\cal C}(\Delta) \geq z$, where $z$ is a random number chosen 
every time step.  Seperate records of $B^*$ and $t_s$ are kept 
for each hemisphere and separate random numbers $z$ are chosen.

After deciding where and when a spot pair should be placed,
the next step is to specify its 2D (latitude,longitude) profile
on the solar surface which we write as
\begin{equation}\label{Bsurf}
B_R(\theta,\phi) = {\cal S} B_s  ~ \left[g_T(\theta,\phi) - g_L(\theta,\phi)\right] 
\end{equation}
where ${\cal S}$ is the sign of $B^*$ at the (co)latitude and longitude of the
spot pair, $\theta_s$ and $\phi_s$. The functions $g_L(\theta,\phi)$ 
and $g_T(\theta,\phi)$ are Gaussian or polynomial profiles defining the leading
and trailing spots. For example, $g_L(\theta,\phi) = 1 - 3s^2 + 2 s^3$ 
for $s \leq 1$ where $s^2 r_s^2 = (\theta - \theta_L)^2 + (\phi - \phi_L)^2$
and $r_s$ is the angular radius of each spot (see below). A similar 
expression holds for $g_T(\theta,\phi)$.

Each spot pair is given a tilt in accordance with Joy's law, as seen in solar 
observations; $\delta = \delta_0 \cos\theta$ where $\delta_0 = 32^\circ.1 \pm 0^\circ.7$ 
\citep{stenf12}. This gives $\theta_{L/T} = \theta_s \pm s_r \sin\delta$ and 
$\phi_{L/T} = \phi_s \pm s_r \cos\delta$. The angular distance between spots, 
$s_r$, is an input parameter (here equal to 3 $r_s$). For an illustration of the
resulting surface field see Fig.\ \ref{fig:spot}$a$.

The field strength $B_s$ and radius $r_s$ of the spot pair are determined
by the flux content
\begin{equation}\label{phispot}
\Phi_s = 2 \Phi_0 ~ \frac{\vert \hat{B}(\theta_s,\phi_s,t)\vert}{B_q} ~ 
\frac{10^{23}}{1 + (\hat{B}(\theta,\phi)/B_q)^2} ~~ \mbox{Mx} = 
B_s r_s^2 ~~~~.
\end{equation}
Here $B_q$ is a quenching field strength (here 10$^5$ G) and $\Phi_0$ is an amplification
factor that can be adjusted to promote supercritical dynamo action.
Solar observations suggest $\Phi_0 \sim 1$, implying a flux of 10$^{23}$ Mx 
in the strongest active regions but for the preliminary proof-of-concept
models presented here, we use fewer, stronger spots,
with $\Phi_0 = 200$, $\tau_p = 400$ days, and $\tau_s = $ 600 days.  
Typically we specify the spot strength as an input parameter 
$B_s =  3000 ~ \Phi_0$ G and set $r_s = (\Phi_s / B_s)^{1/2}$.
However, we often find it practical to set minumum and maximum
values for $r_s$ (here 8-43 Mm), and then adjust $B_s$ accordingly 
to give the desired flux.  For $\Phi_0 = 200$, this preliminary
procedure yields artificially strong spots of 600-1500 kG.  Future 
models will incorporate more realistic spot/BMR distributions.

The 3D structure of the field in a given spot pair is computed
by doing a potential field extrapolation below the surface,
$\BB_{spot}(r,\theta,\phi) = \del \Gamma$ where 
\begin{equation}\label{3dfield}
\Gamma(r,\theta,\phi) = \sum_{\ell m} \left(a_{\ell m} r^\ell + b_{\ell m} r^{-(\ell+1)}\right)
Y_{\ell m}(\theta,\phi) ~~~. 
\end{equation}
The coefficients $a_{\ell m}$ and $b_{\ell m}$ are chosen such that 
$B_r(r,\theta,\phi) = B_R(\theta,\phi)$ at $r = R$ and $B_r = 0$
for $r \leq r_p$, where $r_p$ is an input parameter representing the
initial penetration depth of active regions.  Here we use $r_p = 0.95 R$.

We do not expect the subsurface field structure of actual sunspots to be potential.
However, 
equation (\ref{3dfield}) is easy to implement and it makes close
contact with previous axisymmetric BL solar dynamo models in which the
BL source term is assumed to be confined to the surface layers.  This
is justified by solar observations and modeling efforts that suggest active
regions decouple from their roots within a few days after emergence
\citep{schus05b}, a time short compared to the 11-yr solar activity 
cycle.  The other limit, in which active regions remain anchored to
progenitor fields in the lower CZ and tachocline after emergence, 
will be considered in future work (see also YM13).

\begin{figure}
\centering
\centerline{\epsfig{file=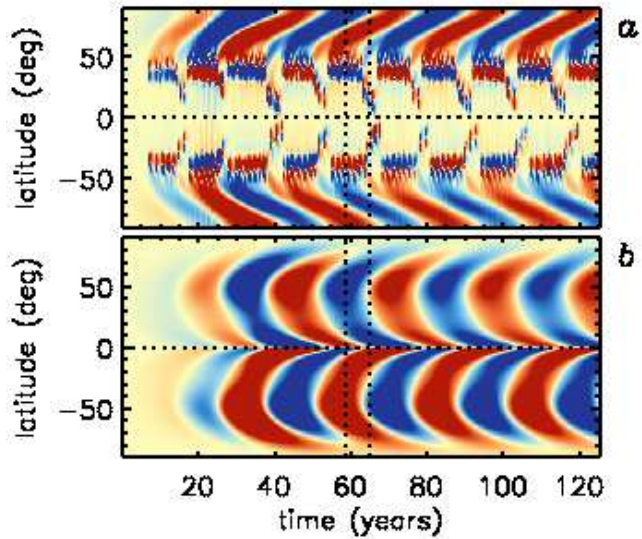,width=\linewidth}}
\caption{\small\label{fig:bfly}
Butterfly diagram for a representative solar dynamo simulation. 
($a$) mean radial field $\left<B_r\right>$ at the surface ($r = R$) 
as a function of latitude and time. Blue 
and red denote inward and outward polarity respectively and the saturation 
level of the color table is $\pm $ 100 G. ($b$) Mean toroidal field
$\left<B_\phi\right>$ near the base of the convection zone ($r = 0.71R$;
blue westward, red eastward, saturation $\pm 30 kG$).  Vertical lines 
denote times of 58.57, 58.83, and 64.94 yr represented in Fig.\ \ref{fig:brss}.  
}
\end{figure}

\section{RESULTS AND DISCUSSIONS}\label{sec:results}


Fig.\ \ref{fig:bfly} highlights magnetic cycles achieved in a solar
dynamo simulation represented in terms of butterfly diagrams.  Shown
are the mean radial field near the surface (Fig.\ \ref{fig:bfly}$a$)
and the mean toroidal field near the base of the CZ (Fig.\
\ref{fig:bfly}$b$) as a function of latitude and time. The numerical
resolution of this simulation is 300$\times$512$\times$1024 in $r$,
$\theta$, $\phi$ (maximum spherical harmonic degree 340) and the
computation domain extends from $0.69R$--$R$, with an electrically
conducting inner boundary and a radial field boundary condition on the
outer surface.

The half-period of the magnetic cycle is roughly 11-12 yrs, comparable
to the solar cycle.  As in other advection-dominated 2D BL/FT models,
this period is regulated largely by the imposed meridional flow
and in particular the equatorward flow of several m s$^{-1}$ near the 
base of the CZ \citep{dikpa09,charb10}.
Still, to our knowlege this is the first published demonstration
of a self-sustained, cyclic solar dynamo model that incorporates 
a 3D flux emergence algorithm for the generation of poloidal field.
The dynamo not only includes sunspots (BMRs), but as a BL model,
it relies on them for its operation.  

\begin{figure}
\centering
\centerline{\epsfig{file=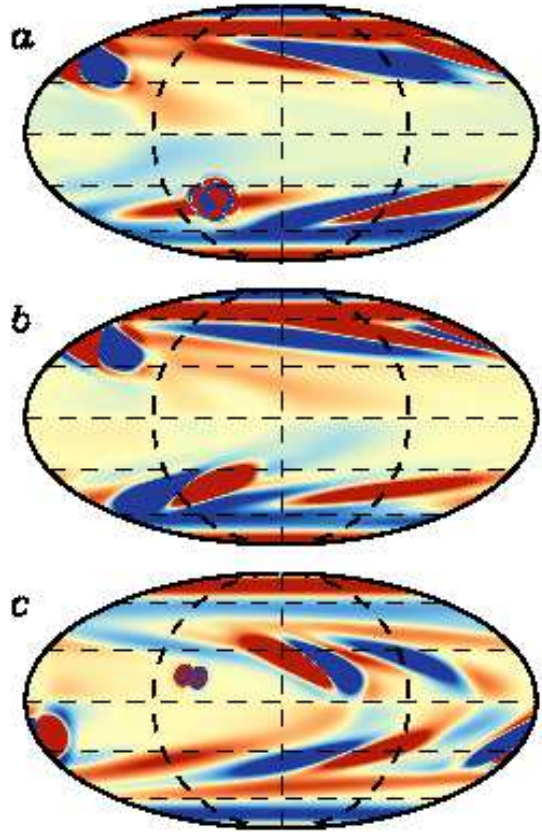,width=\linewidth}}
\caption{\small\label{fig:brss}
Evolution of the radial magnetic field $B_r$ at the solar surface.
Shown are Molleweide projections at the three times indicated by
vertical lines in Fig.\ \ref{fig:bfly}($a$), namely $t = $ 58.57,
58.83, and 64.94 yr. Blue and red denote inward (negative) 
and outward (positive) polarity
with a saturation level of $\pm$ 100 G to highlight relatively weak
fields.  The low saturation level reveals some Gibbs ringing around
the spots as a consequence of the spectral method, but this is two
orders of magnitude smaller than the field at spot center (180 kG and
1.5 MG for the fresh spots in $a$ and $c$ respectively) and
is quickly dissipated, with no significant contribution to the flux
budget.
}
\end{figure}

Fig.\ \ref{fig:brss} shows an illustrative example of surface flux
evolution.  The sequence begins at $t = $ 58.57 yr ($a$) when a new
sunspot pair has just emerged in the SH amid remnant
flux from previous emergence events.  Note also the slighly older spot
pair in the NH at a latitude of about 45$^\circ$ and
longitude near -180$^\circ$.  About three months later ($b$), trailing
flux from the southern spot (blue) has begun to disperse and merge
with a growing axisymmetric band of negative flux at a latitude of
roughly -65$^\circ$.  Similarly, trailing flux from the northern spot
(red) contributes to a positive-polarity band of flux at a latitude of
about 67$^\circ$.  About 6 years later ($c$), these bands (red in the
north and blue in the south) have migrated toward higher latitudes and
have begun to reverse the polar fields (see also Fig.\
\ref{fig:bfly}$a$).  Meanwhile, the next generation of sunspots has
already begun to build opposite-polarity bands equatorward of these.

This surface flux evolution is similar to the evolution of magnetic
flux as seen in photospheric magnetograms and as captured by SFT
models and it demonstrates that the BL mechanism does indeed operate
in a 3D context as originally envisioned by \citet{babco61} and
\citet{leigh64,leigh69}.  The dispersal of tilted
sunspot pairs due to differential rotation, meridional circulation,
and turbulent diffusion generates a mean poloidal field that sustains
the dynamo (see also YM13). Trailing flux from bipolar active regions
migrates toward the poles from mid-latitudes in several streams (Fig.\
\ref{fig:bfly}$a$) while leading flux cancels across the equator.
Note that this cancellation occurs only in a time-integrated sense, 
since the randomness of spot appearances essentially guarantees that 
the 3D field distribution at any instant is not symmetric about the 
equator.

In many previous dynamo models, the mean toroidal field near the base
of the CZ is taken as a proxy for the sunspot number. In our model
this exhibits systematic equatorward propagation at low latitudes
similar to the solar butterfly diagram (Fig.\
\ref{fig:bfly}$b$). However, in our model this proxy is no longer
necessary since we incorporate sunspots (BMRs) explicitly.  Their
behavior in Fig.\ \ref{fig:bfly}$a$ does not agree as well with solar
observations, showing a tendency to linger at mid-latitudes before a
rapid rush toward the equator near the end of a cycle. This can
largely be attributed to the masking function in eq.\ (\ref{Bstar})
which favors mid-latitudes. Since this masking function was originally
designed to mimic Joy's-law tilts that we capture explicitly, it will
be justified to replace it with a more uniform low-latitude profile
that may be calibrated to more closely match solar observations.

As mentioned in \S\ref{sec:spotmaker}, field strengths in this
preliminary model are artificially high due to the prodigious flux
assigned to BMRs.  Typical polar field strengths are 100 G, about an
order of magnitude larger that solar values.  However, the relative
strengths are in reasonable agreement with solar observations in that
most of the magnetic energy (ME) is in the mean toroidal field, which
exceeds the ME in the non-axisymmteric field by a factor of 30-40 and
the ME in the mean poloidal field by a factor of 3000-4000.

In summary, the main result of this letter is the construction of a
viable 3D BL/FT solar dynamo model using a novel Spotmaker algorithm
for flux emergence.  Spotmaker is a 3D generalization of the
double-ring algorithm previously used in 2D BL/FT models and provides
a mechanism for unifying BL/FT dynamo models with SFT models, building
on the sucesses of each.  We focused here on the kinematic regime with
imposed mean flows.  Though this provides an instructive starting
point, the real promise of this model will be realized when we
consider nonlinear feedbacks and non-axisymmetric flows.  By linking
in the full ASH machinery to solve the anelastic equations of motion,
we plan to include Lorentz-force feedbacks and enhanced radiative
cooling in the vicinity of active regions.  This should produce
torsional oscillations as well as modulation of the meridional
circulation and poloidal field generation over the course of multiple
cycles.  We will also consider more realistic sunspot distributions
and alternative flux emergence algorithms such as that proposed by
YM13.  On a longer time scale, we will include resolved convective
motions and investigate their role in the generation and transport of
magnetic flux.

The model presented here is the first step toward a series of
progressively more sophisticated and realistic 3D solar dynamo models
that will allow us to study not only the physics of the dynamo itself, 
but also the response of the corona and heliosphere to cyclic
dynamo-generated fields.

\begin{acknowledgements}
We thank Kyle Augustson and Matthias Rempel for constructrive comments
on the manuscript.
This work is supported by NASA grants NNX08AI57G and NNX08AQ34G. 
The National Center for Atmospheric Research is sponsored by the 
National Science Foundation.
\end{acknowledgements}



\end{document}